\documentclass[pra,showpacs,floatfix,amsmath,amsfonts, twocolumn]{revtex4}
\usepackage{graphicx}% Include figure files
\usepackage[utf8]{inputenc}
\usepackage{dcolumn}% Align table columns on decimal point
\usepackage{bm}% bold math
\usepackage{amsmath}%[fleqn]
\usepackage{amssymb}%[fleqn
\usepackage{wasysym}%[fleqn]
\usepackage{color}

\newcommand{\PT}{{\cal PT}}
\newcommand{\p}{{\cal P}}
\newcommand{\T}{{\cal T}}

\begin{document}

\title{Two-dimensional solitons in conservative and $\mathcal{PT}$-symmetric triple-core waveguides with cubic-quintic nonlinearity}
\author{David Feijoo$^1$, Dmitry A. Zezyulin$^2$ and Vladimir V. Konotop$^2$}
\affiliation{$^1$\'Area de \'Optica, Facultade de Ciencias de Ourense,
Universidade de Vigo, As Lagoas s/n, Ourense, ES-32004 Spain\\
$^2$Centro de F\'isica Te\'orica e Computacional and Departamento de F\'isica,
Faculdade de Ci\^encias da Universidade de Lisboa, Campo Grande, Edif\'icio C8, Lisboa  P-1749-016, Portugal
}

\begin{abstract}
%----------------------------   ABSTRACT  ------------------------------------
We analyze a system of three two-dimensional nonlinear Schr\"odinger equations coupled by linear terms and with the cubic (focusing) -- quintic (defocusing) nonlinearity. We consider two versions of the model: conservative and parity-time ($\PT$) -symmetric ones. These models describe  triple-core nonlinear optical waveguides, with balanced gain and losses in the $\PT$-symmetric case. We obtain families of soliton solutions and discuss their stability. The latter study is performed using a linear stability analysis and checked with direct numerical simulations of the evolutional system of equations. Stable solitons are found in the conservative and $\PT$-symmetric cases.  Interactions and collisions between the conservative and $\PT$-symmetric solitons are briefly investigated, as well.
\end{abstract}
%-----------------------------------------------------------------------------

 %03.75Lm, 42.65.Jx, 42.65.Tg

\maketitle

%--------------------------------------------------------------------------------

\section{Introduction}

The nonlinear Schr\"odinger equation (NLSE) is a canonical model for weakly nonlinear waves in various physical contexts~\cite{Sulem}. In one-dimensional setting, it provides a universal framework for studying bright solitons~\cite{Dodd} existing due to the balance between the dispersion (or diffraction) and the focusing nonlinearity. In two-dimensional (2D) and three-dimensional settings,  bright solitons are unstable and undergo finite-time blowup which manifests itself in a singular growth of the solution amplitude \cite{Sulem}. At the same time, the growing intensity of the wave field makes it necessary to account for  nonlinearities of higher orders, and the collapse can be arrested by the defocusing quintic nonlinearity. This idea has motivated intensive studies of cubic-quintic (CQ) generalizations of NLSE \cite{Quiroga}. Additional relevance of inclusion of CQ nonlinearity into the standard NLSE model is justified by the possibility to establish similarities between propagating light and a liquid for the 2D case \cite{Michinel}. Different characteristics of this ``liquid of light'' were discussed \cite{Paredes}, and its experimental realization was recently reported \cite{Wu}.

Various complex phenomena in nonlinear optics related to the multi-mode propagation can be simulated using models of two coupled NLSEs. In particular, such coupled systems allow one to account for polarization effects~\cite{BF}, describe vector and mixed solitons (i.e., paired bright and dark solitons)~\cite{mixed}, simulate soliton switching~\cite{switching}, and describe the symmetry breaking, the latter corresponding to a transition from a symmetric state which bears identical fields in both components to an asymmetric one~\cite{Malomedcouplings1D,Malomedcouplings}. In the meantime, much less information is available about the wave dynamics in more sophisticated systems of three coupled equations, which, to the best of our knowledge, were mainly studied in the context of the mean-field theory of spinor Bose-Einstein condensates, where  the main attention was focused on the repulsive interactions (i.e., the defocusing nonlinearity in the optical terminology)~\cite{spin-1}.

On the other hand, a natural generalization of coupled NLSE-like systems resides in the possibility of inclusion of the effects related to the gain and loss. One of particularly interesting cases corresponds to the situation of parity-time ($\PT$-) symmetric arrangement of gain and lossy cores \cite{Bender}. The simplest and experimentally feasible $\PT$-symmetric configuration can be implemented in the form of two coupled optical waveguides, one of which experiences gain and another one corresponds to the losses~\cite{Ruter}. Dynamics of solitons in such a $\PT$-symmetric coupler has received  a considerable recent attention in 1D~\cite{Driben} and in 2D~\cite{Malomedgainloss} settings. However, to the best of our knowledge, $\PT$-symmetric solitons in triple-core waveguides have not been reported, so far. In the meantime, it is known that such an important feature as $\PT$-symmetry breaking (i.e., reality of the spectrum of the underlying linear system) is very sensitive not only to the distribution and balance between gain and losses but also to the geometry of the waveguides (depending on whether they are assembled in an open or a closed chain) and to the number of waveguides (either even or odd)~\cite{Barashenkov,Leykam}. This significantly diversifies possible physical scenarios as well as eventual applications of the system.

In the present paper, we address the conservative and $\PT$-symmetric systems of three 2D NLSEs coupled in a circular (closed) chain. More specifically, we classify possible types of vector bright solitons and reveal symmetry breaking bifurcations in the conservative chain. Next, we touch upon the properties of underlying linear problem in the $\PT$-symmetric case where the system possesses a nonzero $\PT$-symmetry breaking threshold, provided that there exists a mismatch in the couplings between the sites. As the main outcome of our work, we numerically show that there exist two branches of $\PT$-symmetric solitons which are stable as long as the strength of the gain-and-loss is small enough. Upon increase of the gain-and-loss parameter the solitons become unstable; however the families of these unstable solutions can be continued to the arbitrary strength of the gain-and-loss, even to the domain of the broken $\PT$ symmetry.

\begin{figure}%[htb]
\includegraphics[width=\columnwidth]{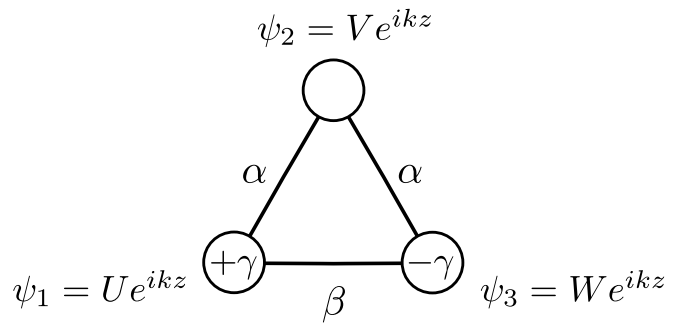}
\caption{A schematic illustration of three coupled equations described by Eqs.~(\ref{3eqsgamma}). A conservative waveguide carrying the field $\psi_2$ is coupled to a $\PT$-symmetric  (for $\gamma>0$) dimer of a gain and lossy waveguides carrying the fields $\psi_1$ and $\psi_3$, respectively. $\alpha$ and $\beta$ are the respective coupling coefficients.}
\label{figtrimer}
\end{figure}

Thus the present work is focused on the model governed by the following equations
\begin{eqnarray}
i \frac{\partial \psi_1}{\partial z} + \nabla^2 \psi_1 + F(|\psi_1|)\psi_1 + \alpha \psi_2 + \beta  \psi_3 &=& i\gamma \psi_1,   \nonumber \label{3eqsgamma} \\
i \frac{\partial \psi_2}{\partial z} + \nabla^2 \psi_2 + F(|\psi_2|)\psi_2 + \alpha \psi_1 + \alpha \psi_3 &=& 0, \quad \quad \quad    \\
i \frac{\partial \psi_3}{\partial z} + \nabla^2 \psi_3 + F(|\psi_3|)\psi_3 + \alpha \psi_2 + \beta \psi_1  &=& - i \gamma \psi_3, \nonumber
\end{eqnarray}
where $\psi_{1,2,3}$ are the dimensionless amplitudes of the electric field in the three cores, $z$ is the propagation distance, $\nabla^2=\frac{\partial^2 }{\partial x^2}+\frac{\partial^2 }{\partial y^2}$ is the 2D Laplace operator in the transverse plane $x$ and $y$, $F(|\psi_j|) = |\psi_j|^2 - |\psi_j|^4$ with $j=1,2,3$ are the CQ nonlinearities, $\alpha>0$ and $\beta>0$ are the coupling coefficients and $\gamma$ is the gain-and-loss parameter. For $\gamma=0$ the system is conservative, as no gain and losses are present. The case $\gamma>0$ preserves the $\PT$ symmetry, where the first equation describes gain, the third equation describes a lossy waveguide, and the second equation remains neutral. From the formal point of view, $\PT$ symmetry manifests itself in the following property: for any solution $\Psi=\left(\psi_{1}(x,y,z),\psi_{2}(x,y,z),\psi_{3}(x,y,z)\right)^T$ where $T$ stands for the transpose   of system (\ref{3eqsgamma}) there also exists another solution $\Psi^{\PT}=\PT\Psi$ where the parity $\p$ is given by 
\begin{eqnarray}
\p=\left(
\begin{array}{ccc}
0 & 0 & 1\\ 0 & 1 & 0 \\ 1 & 0 & 0
\end{array}\right)
 \end{eqnarray}  
 and the anti-linear operator $\T$ acts according to $\T\Psi(x,y,z)=\Psi^*(x,y,-z) $ (hereafter the asterisk $^*$ denotes the complex conjugation).
Notice that $\PT$ symmetry requires not only the balanced gain and loss ($+i\gamma$ in the first waveguide and   $-i\gamma$ in the third waveguide), but also the equal coupling $\alpha$ between the waveguides with  $\psi_1$ and the waveguides with $\psi_2$ and  $\psi_2$ and $\psi_3$. A schematic presentation of the model based on Eqs.~(\ref{3eqsgamma}) is provided in Fig.~\ref{figtrimer}.

After omitting the Laplace operators, system (\ref{3eqsgamma}) is reduced to the $\PT$-symmetric trimer \cite{LiKev} which has been studied before with the cubic nonlinearity \cite{Leykam,trimer}. On the other hand, the introduced system (\ref{3eqsgamma}) can be considered as a generalization of the CQ 2D coupler previously studied both in the conservative \cite{Malomedcouplings} and in the  $\PT$-symmetric \cite{Malomedgainloss} cases.

The remainder of the paper is organized as follows.  In Sec.~\ref{sec:cons}, we explore solitons in the conservative setting, and in Sec.~\ref{sec:PT}  the analysis is extended on the $\PT$-symmetric case.  In Sec.~\ref{sec:coll} we examine interactions and collisions between the solitons. Section~\ref{sec:concl} concludes the paper.

%---------------------------------------------------------------------------------
%%%%%%%%%%%%%%%%%%%%%%%%%%%%%%%%%%%%%%%%%%%%%%%%%%%

\section{Solitons in the conservative model}
\label{sec:cons}

Before considering solitons in the $\PT$-symmetric model, it is of fundamental importance to understand the properties of the underlying conservative model. To this end, in this section we address the case $\gamma=0$ in (\ref{3eqsgamma}).
We start looking for radial stationary soliton solutions of the form:
\begin{eqnarray}
\label{rad_solut}
 \psi_{\{1,2,3\}}  = \{ U(r), V(r), W(r)  \}  e^{i k z},\nonumber \label{3stat}%\\
% \psi_2 = V(r) e^{i k z} \\
% \psi_3 = W(r) e^{i k z}\nonumber
\end{eqnarray}
where $U$, $V$ and $W$ are real functions of the radius $r=\sqrt{x^2+y^2}$ in the $(x,y)$ plane and $k$ is the propagation constant.

The stationary wavefunctions $U$, $V$ and $W$ solve the system
\begin{eqnarray}
-k U + \frac{d^2 U}{d r^2} + \frac{1}{r} \frac{d U}{d r} + F(|U|)U + \alpha V + \beta W &=& 0, \ \ \  \nonumber \label{eqsr}\\
-k V + \frac{d^2 V}{d r^2} + \frac{1}{r} \frac{d V}{d r} + F(|V|)V + \alpha U + \alpha W &=& 0, \ \ \ \\
-k W + \frac{d^2 W}{d r^2} + \frac{1}{r} \frac{d W}{d r} + F(|W|)W + \alpha V + \beta U &=& 0. \ \ \  \nonumber
\end{eqnarray}
The requirement of the regularity of the fields $\psi_{1,2,3}(x,y,z)$ at the origin  $x=y=0$ implies the following boundary condition at $r=0$:  $\left.dU/dr\right|_{r=0}=\left.dV/dr \right|_{r=0} =\left.dW/dr\right|_{r=0}=0$. On the other hand, looking for spatially localized solutions satisfying   $\psi_{1,2,3}(x,y,z)\to 0$ as $x^2+y^2\to\infty$, we require functions $U$, $V$, and $W$ to vanish at the infinity: ${U, V, W}\to 0$ as  $r\to\infty$.
%
%with the boundary conditions $dU/dr(r=0)=dV/dr(r=0)=dW/dr(r=0)=0$ to guarantee the regularity at $r=0$; and $U, V, W \propto \frac{1}{\sqrt{r}} e^{-\Lambda r}$ at $r\to \infty$, where $\Lambda$ can be computed introducing the previous expression in the system (\ref{eqsr}) to obtain $\Lambda^2 = k-\frac{1}{2}\sqrt{\beta^2+8\alpha^2} -\frac{1}{2}\beta$. As it was previously mentioned, our study is focused in the analysis of vector bright solitons, and this last boundary condition proceeds from the fact that the normalized amplitudes $\psi_{\{1,2,3\}}$ must tend to $0$ when $r \to \infty$. %$U(r\to \infty),V(r\to \infty),W(r\to \infty) = 0$

\subsection{Solutions in the limit $\alpha=0$}

\begin{figure}%[htb]
{\centering \includegraphics[width=1.01\columnwidth]{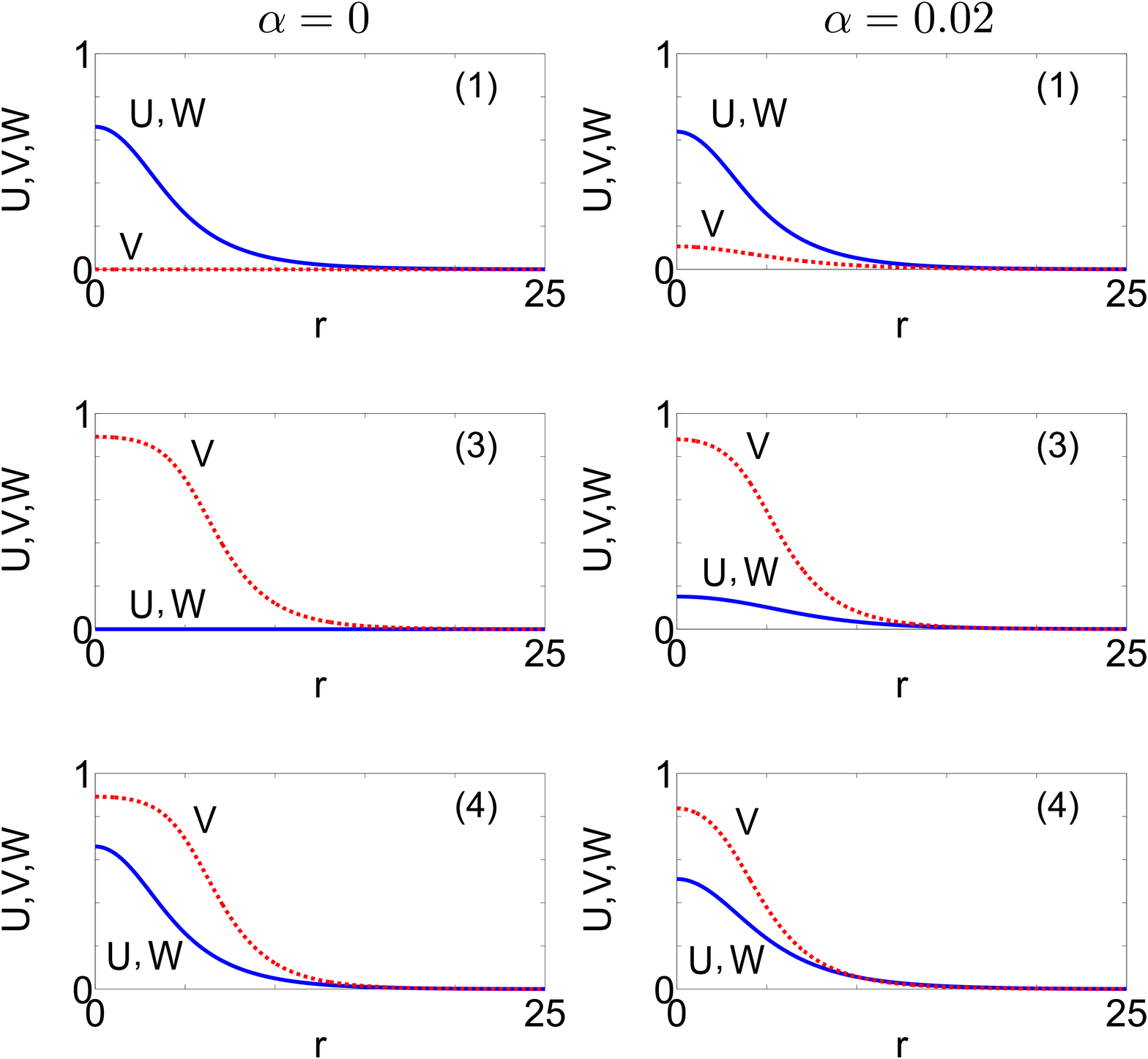}}
\caption{(Color online). Examples of symmetric solutions ($U=W$) of the conservative system found for $k=0.15$ and $\beta=0.07$. The left column presents the profiles for $\alpha=0$ and the right column for $\alpha=0.02$. The solid blue line shows $U$ and $W$, and the dashed red one shows $V$. The number on the right upper side indicates the label of the corresponding symmetric branch.  %For symmetric solutions~1, 2 and 4, one has $U=W$. s
}
\label{fig1}
\end{figure}

\begin{figure}%[htb]
{\centering \includegraphics[width=1.01\columnwidth]{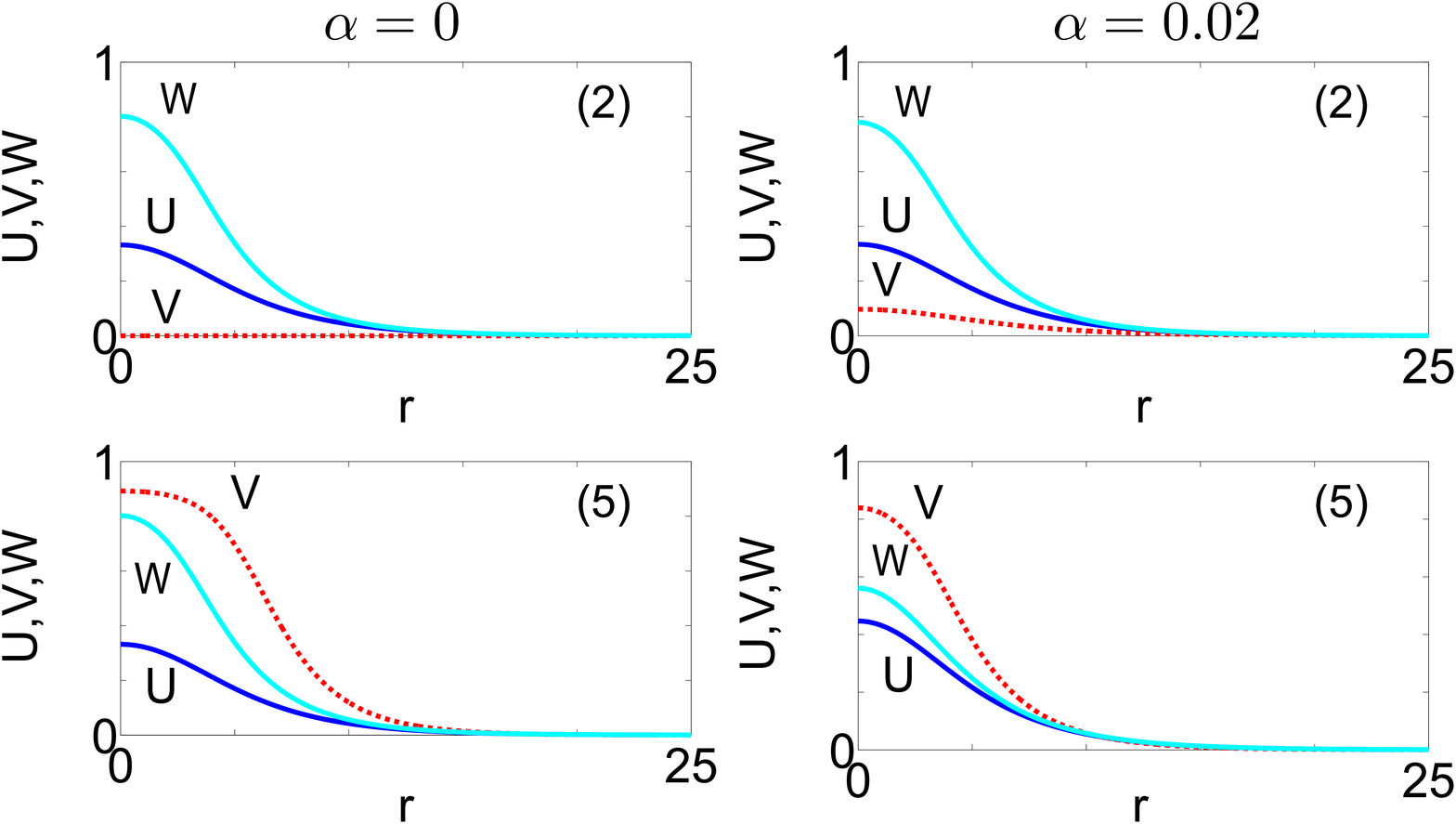}}
\caption{(Color online). Profiles of asymmetric solutions of the conservative coupled waveguides. The values of the parameters are the same that in Fig.~\ref{fig1} and each column shows again the solutions for $\alpha=0$ and $\alpha=0.02$ respectively. The blue and cyan continuous lines correspond to $U$ and $W$, and the dashed red line to $V$. The number on the right upper side indicates the label of the corresponding asymmetric branch.%Each example belongs to one of the five branches obtained from the solutions shown in Fig.~\ref{fig1} by means of continuation in $\alpha$ with constant $k=0.15$, $\beta=0.07$. The solutions are shown for $\alpha=0.02$. As above the blue, cyan continuous and dashed red lines correspond to $U$, $W$, and $V$.
}
\label{fig2}
\end{figure}

In order to classify possible solutions of the system (\ref{eqsr}), it is convenient to start with the limit $\alpha \to 0$ in which system of three equations (\ref{eqsr}) splits into two simpler subsystems whose properties are fairly well understood. The first subsystem is a single CQ-NLSE equation for the wavefunction $V$. It is known that, besides of the trivial zero solution, this equation admits a well-studied solitonic solution \cite{Quiroga, Michinel, Paredes}. The second subsystem consists of two coupled equations for functions $U$ and $W$. Besides of the zero solution, this system admits a branch of \textit{symmetric} solutions, for which $U=W$ and an \textit{asymmetric} branch with $U\ne W$ \cite{Malomedcouplings}. Thus combining the solutions from the two subsystems, we can predict the existence of \textit{five} different nontrivial branches of solutions for the whole system of three equations (\ref{eqsr}) which can be continued to small but nonzero $\alpha$. The solutions of different types can be listed in the following order:
\begin{itemize}
	\item Solution~$1$ is a combination of the symmetric solution for the $U$-$W$ subsystem with the zero solution from the $V$-equation;
	\item Solution~$2$ is a combination of the asymmetric solution for the $U$-$W$ subsystem with the zero solution from the $V$-equation;
	\item Solution~$3$ bears trivial zero solution in the $U$-$W$ subsystem, but the nontrivial solitonic one for the $V$-equation;
	\item Solution~$4$ is the combination of the symmetric solution for the $U$-$V$ coupler with the nonzero solution for the $V$-equation.
	\item Solution~$5$ is the combination of the asymmetric solution for the $U$-$V$ coupler with the nonzero solution for the $V$-equation.
\end{itemize}
These considerations are systematized in the Table~\ref{table} (see the column $\alpha=0$). Examples of the listed solutions are displayed in the left column of Fig.~\ref{fig1} and Fig.~\ref{fig2}.

\begin{table*}
%\centering
%\resizebox{\columnwidth}{!}{%
    \scalebox{0.9}{\begin{tabular}{|c|c|c|c|c|}
        \hline
        Branch No.& symmetry & $\alpha=0$ & arbitrary $\alpha$                                                                                                                                                                            & stability   (for $\alpha >0$)                                                                             \\ \hline
        1            & sym      & $U=W\neq0$, $V=0$       & does not exist for $\alpha>\alpha_{cr}$, see Eq.~(\ref{eq:alpha_cr}) & unstable for small $\alpha$,\\%
        &&&& but becomes  stable after bifurcation with branch 2\\[2mm]
        2             & asym     & $U\neq W$, $V=0$    & does not  exist after the pitchfork bifurcation & stable\\[2mm]
        3              & sym      & $U=W=0$, $V\neq0$       & merges with branch 4&stable  \\ [2mm]
        4              & sym      & $U=W\neq0$, $V\neq0$    & merges with branch 3  & unstable\\[2mm]
  5                   & asym     & $U\neq W$, $V\neq0$ & does not exist after the pitchfork bifurcation & unstable                                                                                \\
        \hline
    \end{tabular}}
\caption{Summary of the main features of the five branches of solutions of the conservative system.}
    \label{table}
\end{table*}

\subsection{Continuation over the coupling parameter $\alpha$}

We use the five solutions identified above in the limit $\alpha \to 0$ as the initial guesses for the numerical continuation over the coupling parameter $\alpha$. % departing from the case $\alpha=0$.
Our numerical results are obtained rewriting the system of equations (\ref{eqsr}) in a finite differences scheme. We introduce a discrete spatial grid in the finite interval $r\in [0, R]$, where $R\gg 1$ is sufficiently large. The zero boundary condition at $r \to\infty$ is approximated by the requirement  $U(R)=V(R)=W(R)= 0$. Given the initial ansatz, solutions are found by a standard Newton-Raphson method (see Ref.~\cite{diff} for details about the finite differences and Newton-Raphson methods).
Each of the five solutions can be continued to nonzero $\alpha$ originating in this way a continuous branch of solutions. The transformation of the soliton shapes under growing $\alpha$  can be traced by comparing the spatial profiles of the solitons in Fig.~\ref{fig1} (symmetric solutions) and Fig.~\ref{fig2} (asymmetric solutions). One observes that for $\alpha>0$ branches~$1$, $3$ and $4$ remain  \textit{symmetric}: i.e., for these branches $U=W$ in the whole range of their existence. Branches~$2$ and $5$ are \textit{asymmetric}, i.e., they do not bear any particular symmetry among the three wavefunctions. Switching on $\alpha$ leads to growth of the second component in the solutions from the  branches~$1$ and $2$ (recall that for the corresponding solutions in the limit $\alpha=0$ the second component vanishes, $V=0$). In a similar way, branch~$3$ has $U=W=0$ for $\alpha=0$, but nonzero $U$ and $W$ ($U=W$) for nonzero $\alpha$.

\begin{figure}%[htb!]
	{\centering {\includegraphics[width=1.05\columnwidth]{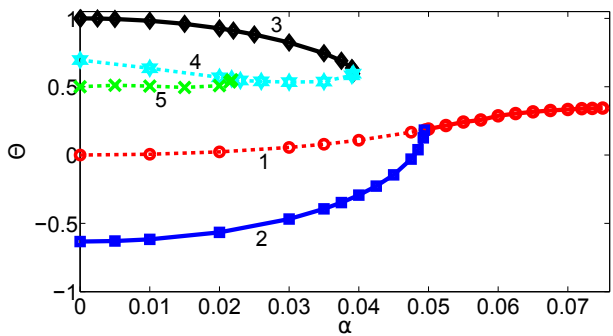}}\\%
		\par}
	\caption{(Color online). Branches of the solutions on the plane $(\Theta,\alpha)$ for $k=0.15$ and $\beta=0.07$. The red circles correspond to solutions of type  $1$, blue squares to type  $2$, black diamonds to type $3$,  cyan hexagons to type  $4$, and green crosses to type  $5$. Dashed lines indicate unstable solutions, and  solid lines show stable solutions.
		%The integrals to calculate the energies (\ref{energies}) in (\ref{Theta}) were cut for all the families to $r_{max}=45$.
	}
	\label{fig3}
\end{figure}

The complete bifurcation diagram obtained numerically after the continuation over the parameter $\alpha$ is visualized in Fig.~\ref{fig3} in the plane $(\Theta,\alpha)$, where the quantity  $\Theta$ is defined as %~(a)
\begin{equation}
\Theta=\frac{E_U+E_V-E_W}{E},
\label{Theta}
\end{equation}
where
\begin{eqnarray}
E_{\{U, V, W\}} =  2\pi \int_{0}^\infty |\{U, V, W\} |^2 rdr, \label{energies}
\end{eqnarray}
are the energies in the corresponding waveguides and
\begin{eqnarray}
 \label{energy_total}
E=E_U+E_V+E_W.
\end{eqnarray}
is the total energy in the system.

The choice of the parameter to characterize bifurcations is not unique. We found $\Theta$ convenient as for the symmetric dimer solutions (the field propagates only in the waveguides $U$ and $V$ at $\alpha=0$) $\Theta=0$ and for the asymmetric solutions it describes the energy imbalance in the dimer. On the other hand, at $\alpha=0$, when energy propagates only along the conservative waveguide, we have $\Theta=1$. Alternatively, a similar diagram plotted in the plane $E$  \textit{vs} $\alpha$ might be thought to be more conventional, but it does not allow to resolve easily the important bifurcation features, since many of the solutions have very close (or virtually equal) energies $E$.
% (e.g. compare the shapes of the solutions in Figs.~\ref{fig1} and ~\ref{fig2}).
Notice also that since the conservative system is invariant under the interchange of $U$ and $W$,
%, i.e., is  symmetric with respect to the vertical axis in the graph in Fig.~\ref{figtrimer},
any asymmetric solution from branches~$2$ and $5$ exists in two ``copies'': $(U, V, W)$ and  $(W, V, U)$ which obviously have different $\Theta$-characteristics. However, since these two copies can be easily obtained one from another, we show only one $\Theta$-dependence for each asymmetric branch, which makes the bifurcation diagram somewhat simpler and easier to read.

The most visible  feature observable in Fig.~\ref{fig3} is that at certain $\alpha$ the asymmetric branch~$2$ (blue squares) merges with the symmetric branch~$1$ (red circles).
%(a), and the latter persists and remains symmetric upon further increase of $\alpha$.
This scenario can be considered as the symmetry breaking through a pitchfork bifurcation. After the bifurcation, symmetric branch~$1$ can be continued until a certain critical value of $\alpha$ at which the solutions lose the exponential localization. The critical value of $\alpha$ can be computed if one looks at the asymptotic behavior of the soliton tails.  Assuming that the behavior of the solutions for large $r$ is given by the following law: $U, V, W \propto \frac{1}{\sqrt{r}} e^{-\Lambda r}$, one can compute 
\begin{equation}
\Lambda^2 =  k-\frac{1}{2}\sqrt{\beta^2+8\alpha^2} -\frac{1}{2}\beta.
\end{equation}
The requirement $\Lambda^2>0$ implies that $\alpha<\alpha_{cr}$, where the critical coupling equals 
\begin{equation}
\label{eq:alpha_cr}
\alpha_{cr} = \frac{1}{\sqrt{2}}\sqrt{k(k-\beta)}.
\end{equation}
It also follows  from (\ref{eq:alpha_cr})  that the   propagation constant  $k$   must be larger than $\beta$:  $k>\beta$.
For the parameters in Fig.~\ref{fig3}, we have $\alpha_{cr} \approx 0.077$. %(a)

The symmetry-breaking scenario in Fig.~\ref{fig3} can also be observed when the asymmetric branch~$5$ (green crosses) meets the symmetric branch~$4$ (cyan  hexagons). After this, the asymmetric branch disappears, and only the symmetric branch~$4$ exists. For larger $\alpha$, the symmetric branch~$4$ merges with the symmetric branch~~$3$ (black diamonds) featuring a saddle-node bifurcation.

\subsection{Stability analysis}

We have also examined the linear stability of the found solutions. Following the standard procedure, we considered perturbed solutions
\begin{eqnarray}
 \psi_1 = e^{i k z} \big[U(r)+U_+(r) e^{i n \theta} e^{\sigma z}+U_-^*(r) e^{-i n \theta} e^{\sigma^* z} \big], \ \  \nonumber \label{3pert}\\
 \psi_2 = e^{i k z} \big[V(r)+V_+(r) e^{i n \theta} e^{\sigma z}+V_-^*(r) e^{-i n \theta} e^{\sigma^* z} \big], \quad \\
 \psi_3 = e^{i k z} \big[W(r)+W_+(r) e^{i n \theta} e^{\sigma z}+W_-^*(r) e^{-i n \theta} e^{\sigma^* z} \big], \nonumber
\end{eqnarray}
where $U_\pm(r)$, $V_\pm(r)$ and  $W_\pm(r)$ describe radial behavior of small perturbations,  $n=0,1, \ldots$ is the azimuthal index of the perturbation, $\theta$ is the polar angle, and $\sigma$ is the eigenvalue whose real part characterizes the instability growth rate. %The symbol $^*$ denotes complex conjugation.
We derived the linear stability eigenvalue problem (see Appendix~\ref{appA}), and computed the instability increment $\max(\textrm{Re}(\sigma))$. We have checked the lowest azimuthal indices with $n=0,1,2,3$, and found that the unstable eigenvalues (if any) are always generated by the perturbation with $n=0$, while the perturbations with $n\geq 1$ do not cause any instability (a similar observation for the system of two equations has been reported in \cite{Malomedcouplings}).

\begin{figure}%[htb!]
{\centering {\includegraphics[width=1.04\columnwidth]{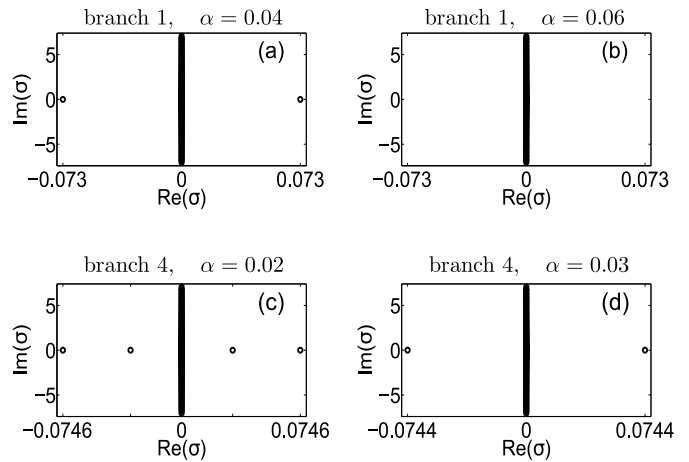}}\par}
\caption{(a,b) The linear stability spectrum for the solution from the symmetric branch~$1$ with $k=0.15$, $\beta=0.07$ and two different $\alpha$.
 %The spectrum contains the one pair of unstable eigenvalues.
 (c,d) The linear stability spectrum for the solutions from the symmetric branch~$4$ with $k=0.15$, $\beta=0.07$ and two different $\alpha$. 
 %The diagram features two pairs of real unstable eigenvalues before the pitchfork bifurcation with asymmetric branch~$5$.
 %The plot shows the existence of only one pair of unstable eigenvalues after the pitchfork bifurcation.%The linear stability spectrum for the solutions from the symmetric branch~$4$ with $k=0.15$, $\beta=0.07$, $\alpha=0.02$ (left panel) and $\alpha=0.03$ (right panel). The first diagram features two pairs of real unstable eigenvalues (before the pitchfork bifurcation with asymmetric branch~$5$), while the second one contains only one pair of unstable eigenvalues (after the pitchfork bifurcation). %The linear stability spectrum for the solution from the symmetric branch~$1$ with $k=0.15$, $\beta=0.07$ and $\alpha=0.04$. The spectrum contains the one pair of unstable eigenvalues.
}
\label{figinst1}
\end{figure}

Linear stability results (also indicated in Fig.~\ref{fig3}) show that in the limit $\alpha=0$ and for small $\alpha$ asymmetric branch~$2$ and symmetric branch~$3$ are stable. The symmetric branch~$1$ is unstable for small $\alpha$ due to a pair of purely real unstable eigenvalues in the stability spectrum [Fig.~\ref{figinst1}(a)], but becomes stable [Fig.~\ref{figinst1}(b)] after the symmetry breaking bifurcation which connects branches $1$ and $2$ (thus the symmetry breaking pitchfork bifurcation connecting branches~$1$ and~~$2$ can be characterized  as supercritical with respect to the parameter $1/\alpha$).  Symmetric branch~$4$ is  unstable in the whole range of its existence. The instability is caused by two pairs of real unstable eigenvalues before the symmetry-breaking bifurcation with asymmetric branch~$5$ [Fig.~\ref{figinst1}(c)]; after the bifurcation, branch~$4$ is unstable due to only one pair of real eigenvalues [Fig.~\ref{figinst1}(d)]. Branch~$5$ has a stable solution with $\alpha=0$ but becomes unstable (with one pair of real eigenvalues) for any nonzero $\alpha$.

%\begin{figure}%[htb!]
%{\centering {\includegraphics[width=0.995\columnwidth]{figure5.eps}}\par}
%\caption{(a) The linear stability spectrum for the solution from the symmetric branch~$1$ with $k=0.15$, $\beta=0.07$ and %$\alpha=0.04$. The spectrum contains the one pair of unstable eigenvalues. (b) The linear stability spectrum for the %solutions from the symmetric branch~$4$ with $k=0.15$, $\beta=0.07$ and $\alpha=0.02$. The diagram features two pairs of %real unstable eigenvalues before the pitchfork bifurcation with asymmetric branch~$5$. (c) The linear stability spectrum %for the solutions from the symmetric branch~$4$ with $k=0.15$, $\beta=0.07$ and $\alpha=0.03$. The plot shows the %existence of only one pair of unstable eigenvalues after the pitchfork bifurcation.%The linear stability spectrum for the %solutions from the symmetric branch~$4$ with $k=0.15$, $\beta=0.07$, $\alpha=0.02$ (left panel) and $\alpha=0.03$ (right %panel). The first diagram features two pairs of real unstable eigenvalues (before the pitchfork bifurcation with %asymmetric branch~$5$), while the second one contains only one pair of unstable eigenvalues (after the pitchfork %bifurcation). %The linear stability spectrum for the solution from the symmetric branch~$1$ with $k=0.15$, $\beta=0.07$ %and $\alpha=0.04$. The spectrum contains the one pair of unstable eigenvalues.
%}
%\label{figinst1}
%\end{figure}

%\begin{figure}%[htb!]
%{\centering {\includegraphics[width=0.49\columnwidth]{figure6a.eps}}%
%{\includegraphics[width=0.49\columnwidth]{figure6b.eps}}\par}
%\caption{The linear stability spectrum for the solutions from the symmetric branch~$4$ with $k=0.15$, $\beta=0.07$, $\alpha=0.02$ (left panel) and $\alpha=0.03$ (right panel). The first diagram features two pairs of real unstable eigenvalues (before the pitchfork bifurcation with asymmetric branch~$5$), while the second one contains only one pair of unstable eigenvalues (after the pitchfork bifurcation).
%}
%\label{figinst4}
%\end{figure}
The stability of the solutions was also checked by means of the direct propagation of the stationary solitons. The input stationary profiles were slightly perturbed as % by asymmetric perturbations:
\begin{eqnarray}
\psi_1(x,y,0) &=& U(r)\,(1 - \epsilon), \quad  \label{solpert}\nonumber \\
\psi_2(x,y,0) &=& V(r)\,(1 + 0.2\epsilon),   \\
\psi_3(x,y,0) &=& W(r)\,(1 + 0.8\epsilon),  \nonumber
\end{eqnarray}
where for numerical simulations we set $\epsilon=0.03$.
The propagation of the perturbed solutions was simulated by means of a split-step pseudo-spectral method, specifically the so-called Beam Propagation Method (BPM) \cite{bpm1}  in a lattice of 512$\times$512 points. This explicit method is conditionally stable, so that a sufficiently small step $\Delta z$ must be considered  \cite{bpm2}. Even though  the scheme is of the first order in $\Delta z$, the evolution associated to the non-derivative terms was computed with a fourth order Runge-Kutta method. The perturbed solutions were propagated to a long distance ($z \gtrsim  600$) to observe their evolution.
\begin{figure}%[htb!]
{\centering {\includegraphics[width=78mm]{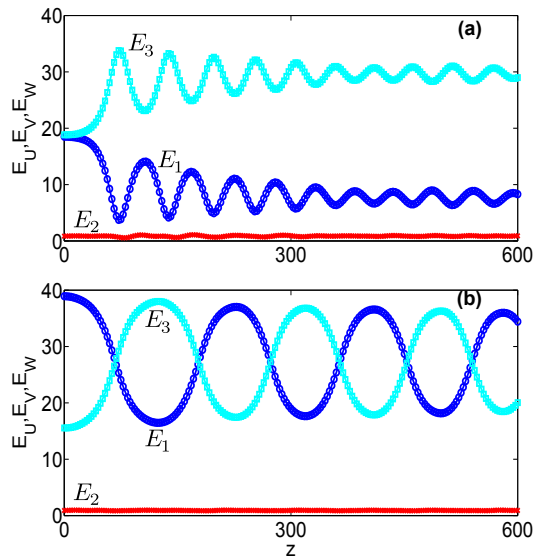}}\par} %0.90\columnwidth
\caption{(Color online). (a) Plot of the energies in each waveguide $E_{1,2,3}=\int_{\mathbb{R}^2} |\psi_{1,2,3}|^2dxdy$ \textit{vs.} propagation distance $z$ for the unstable symmetric solution from branch~$1$ with $k=0.15$, $\beta=0.07$ and $\alpha=0.02$. (b) Plot of the energies   $E_{1,2,3}$ \textit{vs.} propagation distance $z$ for the unstable asymmetric solution from branch~$2$ with $k=0.174$, $\beta=0.07$ and $\alpha=0.02$. While the $z$-axis of panel (b) is limited to $z=600$, we have checked that the shown regular oscillations of the energies persist   until  {$z=3500$}. For larger propagation distances (not shown in the figure), the solution loses spatial localization.
%In both panels, the blue circles and cyan squares correspond to profiles $U$ and $W$, and red crosses to profile $V$.
}
\label{figunst1}
\end{figure}
%\begin{figure}%[htb!]
%{\centering {\includegraphics[width=0.95\columnwidth]{figure6.eps}}\par}
%%{\centering {\includegraphics[width=\columnwidth]{figure6a.eps}}\\%
%%{\includegraphics[width=\columnwidth]{figure6b.eps}}\par}
%\caption{(Color online). (a) Plot of the energies of the profiles $E_U$, $E_V$ and $E_W$ \textit{vs.} propagation distance $z$ for the unstable symmetric solution from branch~$1$ with $k=0.15$, $\beta=0.07$ and $\alpha=0.02$. (b) Plot of the energies of the profiles $E_U$, $E_V$ and $E_W$ \textit{vs.} propagation distance $z$ for the unstable asymmetric solution from branch~$2$ with $k=0.174$, $\beta=0.07$ and $\alpha=0.02$. While the $z$-axis of panel (b) is limited to $z=600$, we have checked that the shown regular oscillations of the energies persist   until \textcolor{red}{$z=3500$}. For larger propagation distances (not shown in the figure), the solution loses spatial localization.  In both panels, the blue circles and cyan squares correspond to profiles $U$ and $W$, and red crosses to profile $V$.
%}
%\label{figunst1}
%\end{figure}
The  results obtained from the simulation of the beam propagations agree with the above conclusions on the linear stability analysis. \\
For stable solutions, amplitude of the perturbation does not grow. For unstable solutions, the growing perturbation destroys the solutions which eventually become non-localized and lose completely their original shape. However, unstable solutions from branches $1$, $2$ and $3$ can preserve localization for significantly long propagation distance. During this long transient period, the instability manifests itself in almost periodic power oscillations whose amplitude decreases slowly.  An example of such a behavior for an initially symmetric unstable solution from branch $1$ is illustrated in  Fig.~\ref{figunst1}(a) and  Fig.~\ref{figunst2}.  As shown in Fig.~\ref{figunst2}, the initially symmetric solution develops strong asymmetry. Fig.~\ref{figunst1}(b) and Fig.~\ref{figunst3} illustrate the development of almost periodic oscillations for an unstable asymmetric mode from branch 2.
\begin{figure}%[htb]
{\includegraphics[width=1.025\columnwidth]{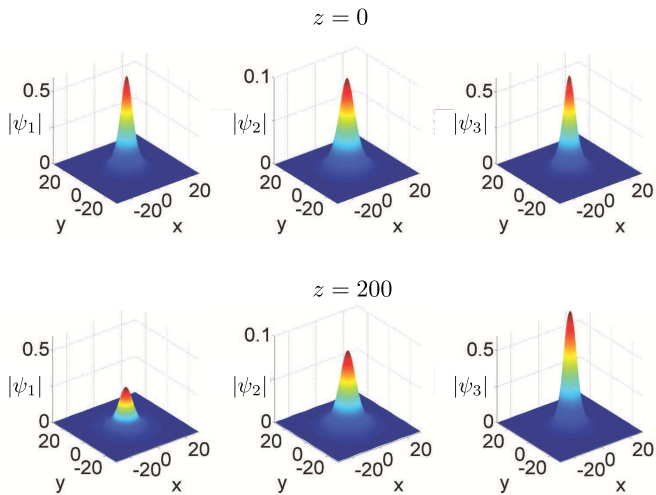}\par}
\caption{(Color online).  Snapshots of the unstable initially symmetric solution whose propagation is shown in  Fig.~\ref{figunst1}(a).  The plots of the first file show the initial solution, and the plots of the second one correspond to $z=200$.
}
\label{figunst2}
\end{figure}

%\begin{figure}%[htb]
%{\includegraphics[width=\columnwidth]{figure8.eps}\par}
%	%\includegraphics[width=0.22\textwidth]{figure9a.eps}
%	%\includegraphics[width=0.22\textwidth]{figure9b.eps}\\
%	%\includegraphics[width=0.22\textwidth]{figure9c.eps}
%	%\includegraphics[width=0.22\textwidth]{figure9d.eps}\\
%	%\includegraphics[width=0.22\textwidth]{figure9e.eps}
%	%\includegraphics[width=0.22\textwidth]{figure9f.eps}
%	\caption{(Color online). Snapshots of the unstable solution whose propagation is shown in Fig.~\ref{figunst1}(b). The plots of first file show the initial solution, and the plots of the second one correspond to $z=300$.
%	}
%	\label{figunst3}
%\end{figure}
\begin{figure}%[htb]
{\includegraphics[width=1.025\columnwidth]{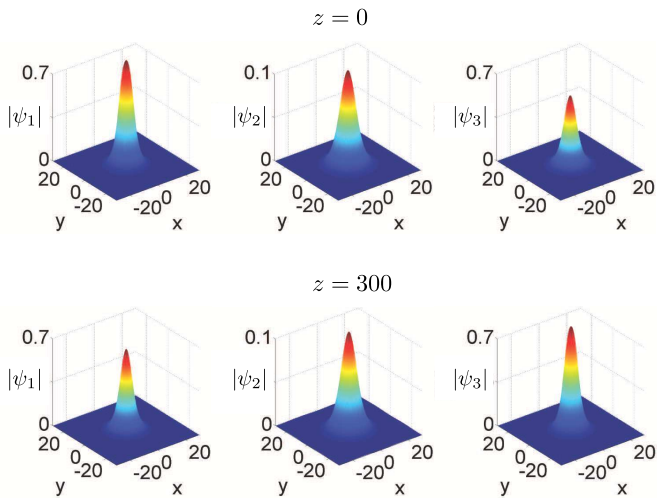}\par}
	\caption{(Color online). Snapshots of the unstable solution whose propagation is shown in Fig.~\ref{figunst1}(b). The plots of first file show the initial solution, and the plots of the second one correspond to $z=300$.
	}
	\label{figunst3}
\end{figure}

\subsection{Families of solutions: continuation over the propagation constant $k$}

\begin{figure}%[htb]
	{\includegraphics[width=1.05\columnwidth]{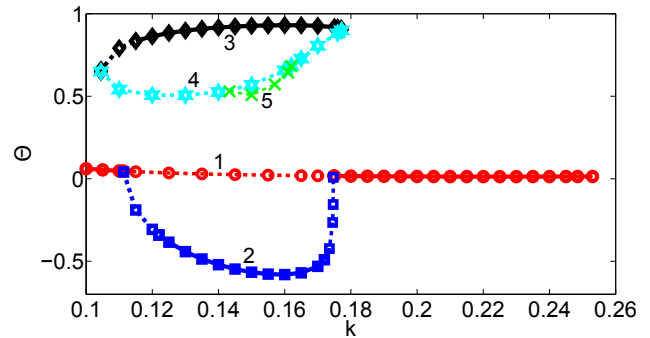}\par}
	\caption{(Color online). Families of solutions on the plane  $(\Theta,k)$ for $\beta=0.07$ and $\alpha=0.02$. In both panels, the red circles correspond to solutions of type  $1$, blue squares to type  $2$, black diamonds to type $3$,  cyan hexagons to type  $4$, and green crosses to type  $5$. Dashed lines indicate unstable solutions, and  solid lines show stable solutions.
	}
	\label{figk}
\end{figure}

As the next step, we constructed families of the solutions $\Theta(k)$ (continuing solutions of the stationary problem over the propagation constant $k$ with all other parameters fixed). The obtained families are visualized in Fig.~\ref{figk} on the plane $(\Theta,k)$. Similarly to what has been observed in \cite{Malomedcouplings} for a coupler, we obtain that the possible values of the propagation constant belong to the range from
$k_{min}=\beta$ up to $k_{max}=\beta+3/16$, where $k_0=3/16$ is the maximal value in the single 2D CQ-NLSE model \cite{Prytula} in view of the divergence of the total energy $E$.

The bifurcation diagram in Fig.~\ref{figk} also features the symmetry breaking where the asymmetric family~$2$ (blue squares) branches off from the symmetric family $1$ via a pitchfork bifurcation. The diagram also shows the exchange of stability which takes place after the bifurcation.

\section{$\PT$-symmetric solitons}
\label{sec:PT}

\subsection{$\PT$ symmetry breaking in the linear model}

Before proceeding to the solitonic solutions in the nonlinear $\PT$-symmetric model (\ref{3eqsgamma})  with $\gamma >0$, we recall the features of $\PT$-symmetry breaking in the underlying linear model.  Omitting for the time being the CQ nonlinear part, we make the Fourier transform of the resulting linear model. Introducing $\hat{\psi}_j = \int_{-\infty}^\infty\int_{-\infty}^\infty   e^{ik_xx + ik_yy}\psi_j(x,y,z)dxdy$, and the column vector $\hat\psi=(\hat{\psi}_1,\hat{\psi}_2,\hat{\psi}_3)^T$, where $T$ stands for the transpose, we obtain
\begin{eqnarray}
 i \frac{\partial\hat{\psi}}{\partial z} =(k_x^2+k_y^2)\hat{\psi}- H\hat{\psi}
\end{eqnarray}
where
\begin{eqnarray}
  \quad
H=  \begin{pmatrix}-i \gamma & \alpha & \beta \\\alpha & 0 & \alpha \\\beta & \alpha & i \gamma \end{pmatrix} .
\label{matriz}
\end{eqnarray}
\normalsize

The linear waves of the system are stable (and hence $\PT$ symmetry is unbroken) if all the eigenvalues of the matrix $H$ are real. The spectrum of (\ref{matriz}) can be easily found~\cite{Leykam}. In particular the condition of the unbroken $\PT$ symmetry reads
\begin{equation}
\gamma^2 \leq \gamma_\PT^2 =  2\alpha^2 + \beta^2 - 3\sqrt[3]{\alpha^4\beta^2}.
\label{PTcondition}
\end{equation}
%Notice that $\Delta$ does not depend on $K$, which is natural to expect.
Condition (\ref{PTcondition}) implies that for any $\alpha$ and $\beta$ there exists a $\PT$-symmetry breaking threshold $\gamma_\PT$ such that $\PT$ symmetry is unbroken if $0\leq \gamma \leq \gamma_\PT$, but becomes broken otherwise.
%For a specific value of $\alpha$ and $\beta$, solving Eq.(\ref{PTcondition}) is possible to obtain the allowed range of values for $\gamma$ which accomplish the unbroken $\PT$ symmetry.
For the case of the homogeneous coupling, i.e., for  $\alpha=\beta$ one has $\gamma_\PT=0$ \cite{LiKev}, that is $\PT$ symmetry is broken for any nonzero gain-and-loss parameter $\gamma>0$.

\subsection{Solitons}
The system (\ref{3eqsgamma}) admits stationary $\PT$-symmetric solitons in the form (\ref{rad_solut})
%\begin{eqnarray}
%\label{3stat2}
% \psi_1 = U(r) e^{i k z},  \,\,\,
% \psi_2 = V(r) e^{i k z}, \,\,\,
% \psi_3 = W(r) e^{i k z},
%\end{eqnarray}
where  for $\gamma>0$ we assume that $U(r)=W^*(r)$ is, generically speaking, complex-valued function, and $V(r)$ is a real-valued function. Substituting (\ref{rad_solut}) in (\ref{3eqsgamma})  and separating the wavefunction $U$ into real and imaginary parts,  $U(r)=U_{\textrm{R}}(r)+iU_{\textrm{I}}(r)$, we arrive at the following system
\begin{eqnarray}
 \frac{d^2 U_{\textrm{R}}}{d r^2} + \frac{1}{r} \frac{d U_{\textrm{R}}}{d r} + [ U_{\textrm{R}}^2 + U_{\textrm{I}}^2 - (U_{\textrm{R}}^2 + U_{\textrm{I}}^2)^2]U_{\textrm{R}} -k U_{\textrm{R}} \label{eqsr2} \nonumber
 \\
 + \alpha V + \beta U_{\textrm{R}} + \gamma U_{\textrm{I}} = 0,
 \nonumber \\[2mm]
 \frac{d^2 V}{d r^2} + \frac{1}{r} \frac{d V}{d r} + (V^2 - V^4)V + 2\alpha U_{\textrm{R}}-k V = 0, \  \
  \\[2mm]
 \frac{d^2 U_{\textrm{I}}}{d r^2} + \frac{1}{r} \frac{d U_{\textrm{I}}}{d r} + [ U_{\textrm{R}}^2 + U_{\textrm{I}}^2 - (U_{\textrm{R}}^2 + U_{\textrm{I}}^2)^2]U_{\textrm{I}} -k U_{\textrm{I}} \quad
 \nonumber  \\
 - \beta U_{\textrm{I}} - \gamma U_{\textrm{R}} = 0.
\nonumber
\end{eqnarray}

In order to obtain $\PT$-symmetric solitons with $\gamma>0$, we use the numerical continuation from the conservative limit $\gamma=0$. As follows from the requirement $U(r) = W^*(r)$, we can calculate $\PT$-symmetric solitons starting only from the symmetric conservative solutions, i.e., from the solutions of branches $1$, $3$ and $4$.  We have also checked the stability of all $\PT$-symmetric solutions following the previous linear stability analysis with the incorporation in the system of equations (\ref{3pertsolv}) the terms responsible for gain and loss.

\begin{figure}%[htb]
{\includegraphics[width=1.05\columnwidth]{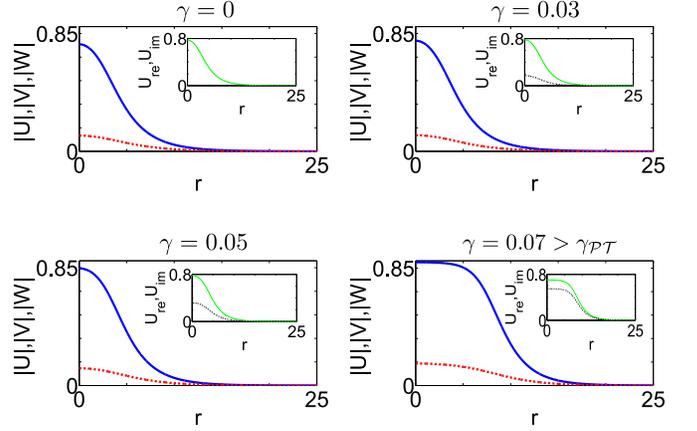}\par}
\caption{(Color online) Radial profiles of 2D $\PT$-symmetric solitons for the branch~$1$ with $k=0.18$, $\beta=0.07$,  $\alpha=0.02$ and   different $\gamma$. For the chosen parameters, the $\PT$-symmetry breaking threshold equals to $\gamma_{\PT}\approx 0.054$.   The solid blue line shows $|U|=|W|$, and the dashed red line shows $|V|$. In the insets, the real (solid green line) and imaginary (dashed black line) parts of $U$ are displayed.
}
\label{fig4}
\end{figure}

\begin{figure}%[htb]
{\includegraphics[width=1.05\columnwidth]{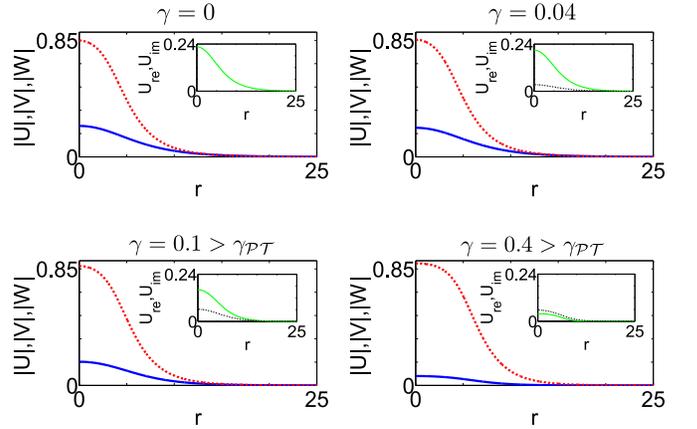}\par}
\caption{(Color online). Radial profiles of 2D PT-symmetric solitons for the branch~$3$ with $k=0.15$, $\beta=0.07$,  $\alpha=0.03$ and different values of $\gamma$. For the chosen parameters, the $\PT$-symmetry breaking threshold equals to $\gamma_{\PT}\approx 0.044$.  The solid blue line shows $|U|=|W|$, and the dashed red line shows $|V|$. The insets show the profiles for the real (solid green line) and imaginary (dashed black line) parts of $U$.
}
\label{fig5}
\end{figure}

As discussed above, all conservative solutions from branch~$4$ are unstable.  We have observed that $\PT$-symmetric solutions obtained  from this branch by considering $\gamma>0$ remain unstable. Therefore, in what follows  we focus on branches $1$ and $3$ which  can  generate stable $\PT$-symmetric solitons.  Figure~\ref{fig4} and Fig.~\ref{fig5} display examples of numerically found $\PT$-symmetric solutions obtained by means of the continuation from branches~$1$ and $3$. Notice a well-pronounced difference between the two types of solitons: for the solitons in Fig.~\ref{fig4}, the amplitude of the cores with gain and losses is larger than the amplitude of the neutral core, i.e., $|U|=|W|>|V|$; while for the solitons in Fig.~\ref{fig5} we have $|V|>|U|=|W|$. In Fig.~\ref{fig4}, the increase of $\gamma$ leads to the progressive increase of $|U|=|W|$, whereas in Fig.~\ref{fig5} the opposite takes place: the amplitudes $|U|=|W|$ decrease as the gain-and-loss parameter $\gamma$ increases.

\begin{figure}%[htb]
{\includegraphics[width=\columnwidth]{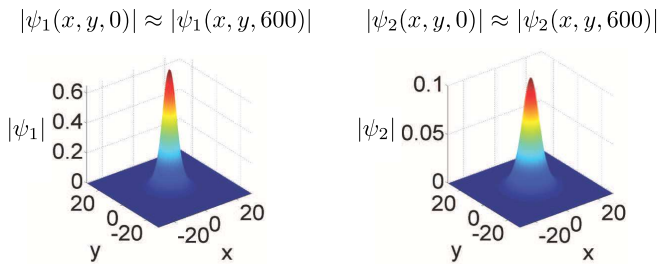}\par}
\caption{(Color online). Long-distance evolution of a stable 2D $\PT$-symmetric soliton from the branch~$1$. The values of the parameters are the same as in Fig.~{\ref{fig4}} except for $\gamma=0.025$. The slightly perturbed initial beam propagates  undistorted at least until  $z=600$. The profile  of $|\psi_3|$ is  almost identical to $|\psi_1|$ and therefore is not shown.
}
\label{fig6}
\end{figure}

\begin{figure}%[htb]
{\includegraphics[width=1\columnwidth]{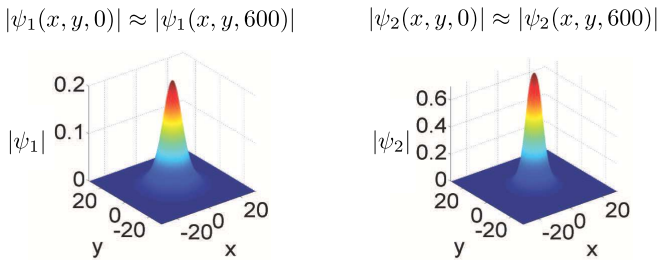}\par}
\caption{(Color online). Long-distance evolution of a stable 2D $\PT$-symmetric soliton from the branch~$3$. The values of the parameters are the same as in Fig.~{\ref{fig5}} except for  $\gamma=0.02$. The slightly perturbed initial beam propagates practically undistorted at least until  $z=600$. The profile  of $|\psi_3|$ is  almost identical to $|\psi_1|$ and therefore is not shown.}
\label{fig7}
\end{figure}

%The stability of the $\PT$-symmetric solitons was studied using the linear stability analysis and the results were checked with numerical simulations employing perturbed solutions in the form of (\ref{solpert}). %allow to conjecture
Our numerical results show that stable conservative solitons (with $\gamma=0$) give birth to $\PT$-symmetric solitons (with $\gamma>0$) which remain stable, at least for sufficiently small $\gamma$. Moreover, our results allow to conjecture that the $\PT$-symmetric solitons continued from a stable conservative soliton remain stable for \emph{any} $\gamma$ below  the $\PT$-symmetry breaking threshold  (we however notice that an accurate analytical treatment  is required in order to substantiate this conjecture; in the vicinity of the  $\PT$-symmetry breaking, i.e., for  $0<\gamma_{\PT}-\gamma\ll 1$  hypothetical instability can be present, but its increment is   small (of order $10^{-3}$ or less), and the associated eigenfunction is poorly localized, which requires a nonpractically large spatial window in order to perform an accurate computation).  Stable solutions propagate for indefinitely long distance without the growth of the initially introduced perturbation. Figures~{\ref{fig6}} and  {\ref{fig7}}  show two representative examples, where the shape of a slightly perturbed  initial beam is practically indistinguishable from the beam obtained after the long-distance propagation.


The families of $\PT$-symmetric solitons can be numerically continued to arbitrarily large values of the gain-and-loss parameter $\gamma$ and, in particular, to the domain of the broken $\PT$ symmetry, i.e., to $\gamma>\gamma_\PT$, as this is typically occur in the stationary oligomer models~\cite{Leykam,LiKev,trimer}.
However, all solitons with  $\gamma > \gamma_{\PT}$  are   unstable. Examples of such unstable solitons are shown in the two   panels of Fig.~\ref{fig4} and Fig.~\ref{fig5} labeled as $\gamma>\gamma_\PT$ . Finally, we point out a difference between our system and the  $\PT$-symmetric system of two equations \cite{Malomedgainloss}: in the latter one no $\PT$ symmetric solitons (either stable or unstable) exists for $\gamma>\gamma_\PT$.

%\begin{figure}%[htb]
%{\includegraphics[width=0.7\columnwidth]{figure14.eps}\par}
%%\includegraphics[width=0.22\textwidth]{figure14a.eps}
%%\includegraphics[width=0.22\textwidth]{figure14b.eps}\\
%%\includegraphics[width=0.22\textwidth]{figure14c.eps}
%%\includegraphics[width=0.22\textwidth]{figure14d.eps}\\
%\caption{(Color online). Evolution of a weakly unstable solution with $\gamma=0.045$ which is slightly less than the %$\PT$-symmetry breaking threshold  $\gamma_\PT$. Values of the other parameters are the same as in Fig.~{\ref{fig6}}.
%}
%\label{fig8}
%\end{figure}
%\begin{figure} %[h]
%{\includegraphics[width=0.7\columnwidth]{figure15.eps}\par}
%%\includegraphics[width=0.22\textwidth]{figure15a.eps}
%%\includegraphics[width=0.22\textwidth]{figure15b.eps}\\
%%\includegraphics[width=0.22\textwidth]{figure15c.eps}
%%\includegraphics[width=0.22\textwidth]{figure15d.eps}\\
%\caption{(Color online). Evolution of a weakly unstable soliton with $\gamma=0.044$ which is slightly less than the %$\PT$-symmetry breaking threshold $\gamma_\PT$. Values of the other parameters are the same as in Fig.~{\ref{fig8}}. %It %can be observed how the central peaks remain almost undistorted and the emitted radiation becomes only well visible on %the soliton tails.
%}
%\label{fig9}
%\end{figure}

\section{A note on interaction between solitons}
\label{sec:coll}
\begin{figure}%[htb]
{\centering {\includegraphics[width=0.9\columnwidth]{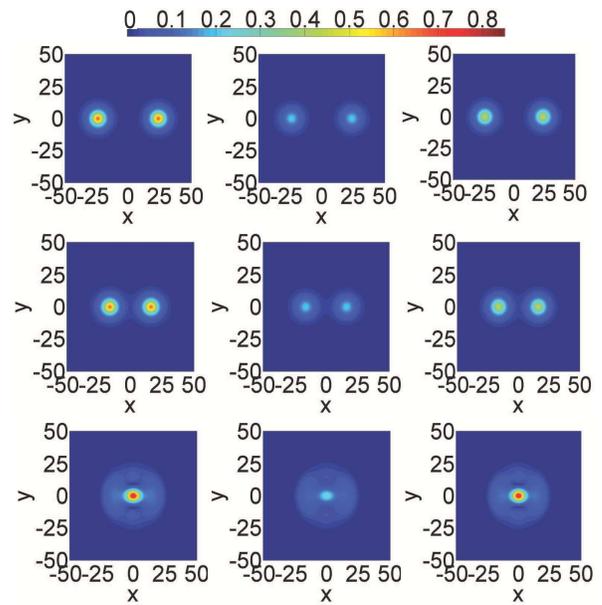}}\\%
		\par}
\caption{(Color online). Example of interaction between asymmetric solitons of the branch~$2$ with $k=0.16$, $\beta=0.08$ and $\alpha=0.04$. The initial distance between them is $\Delta x=46$ and the relative phase $\Delta\phi=0$. The first, the second and the third columns visualize behavior of $\psi_1$, $\psi_2$ and $\psi_3$, respectively. The first, the second and the third rows correspond to  $z=0$, $z=2300$ and $z=2558$, respectively.
}
\label{fig10}
\end{figure}

In this section, we present a brief study on the interactions and collisions of the solitons described by (\ref{3eqsgamma}). First, we analyzed interaction of asymmetric solitons in the conservative system. Initially quiescent solitons had a relative phase $\Delta\phi$ between them and were separated by a relatively small distance $\Delta x$. The respective initial distributions were prepared as follows:
\begin{eqnarray}
\{U,V,W\}_{in}(x,y)= \{U,V,W\}_{st}(x-\Delta x,y)
\label{eqsint} \nonumber
\\+\{U,V,W\}_{st}(x+\Delta x,y)e^{i \Delta \phi}.
\end{eqnarray}
We have observed that in-phase solitons, $\Delta\phi=0$, attract each other and undergo inelastic collision after which they merge in a single pulse, as shown in Fig.~\ref{fig10}. Out-of-phase solitons, $\Delta\phi=\pi$, repel each other, similarly to what happens in the system of  two equations  \cite{Malomedcouplings}.

 We have also considered two $\PT$-symmetric solitons which were launched towards each other with an initial velocity from a certain distance. After the collision, the solitons combine one more time in a single pulse.

\begin{figure}%[htb]
{\centering {\includegraphics[width=53mm]{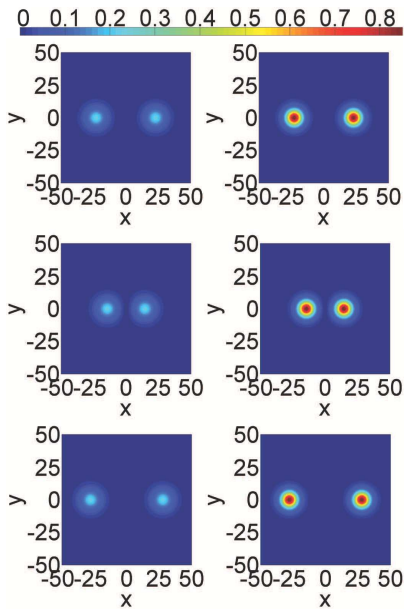}}\\%[width=58mm][width=0.85\columnwidth]
		\par}
\caption{(Color online). Collision between $\PT$-symmetric solitons of the branch~$3$ with $k=0.15$, $\beta=0.07$, $\alpha=0.03$ and $\gamma=0.02$. The initial distance between them is $\Delta x=46$, the relative phase $\Delta\phi=\pi$ and the velocity $v=0.02$. The first column shows $|\psi_1|=|\psi_3|$, and the second column visualizes the amplitude of the second component $\psi_2$. The first, the second and the third rows correspond to $z=0$, $z=264$ and $z=816$, respectively.
}
\label{fig11}
\end{figure}

\begin{figure}%[htb]
{\centering {\includegraphics[width=0.95\columnwidth]{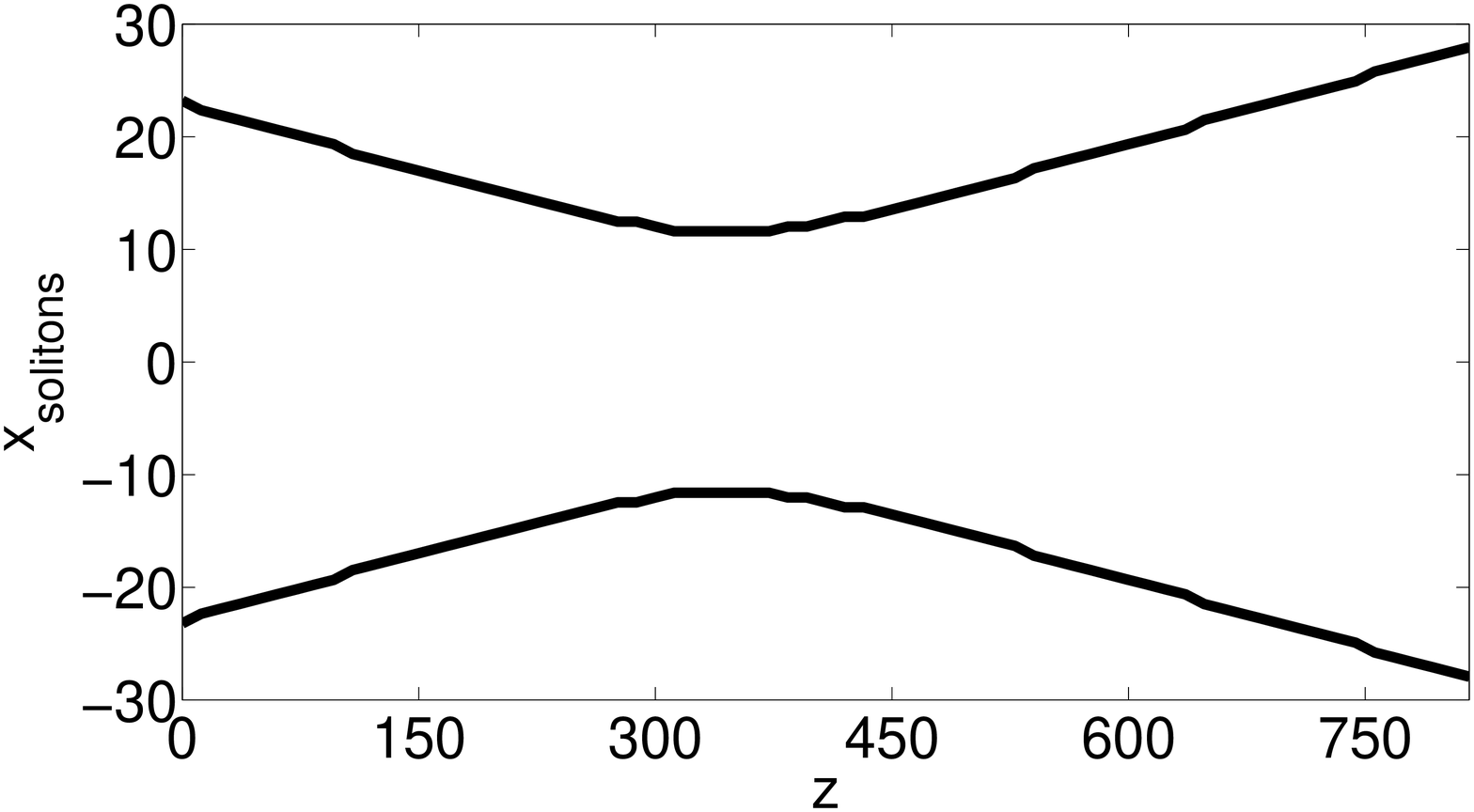}}\\%
		\par}
\caption{(Color online). Estimated positions in $x$ of the $\PT$-symmetric solitons of Fig.~{\ref{fig11}} \textit{vs.} the propagation distance $z$. The continuous black lines indicate the positions of the solitons' peaks. It can be appreciated how the repulsion prevents the cross of solitons at $x=0$.
 %the closer between the solitons.%Estimated positions in $x$ of the $\PT$-symmetric solitons of Fig.~{\ref{fig11}} for a specific value of $|\psi|$ \textit{vs.} the propagation distance $z$. The continuous black line indicates the positions for $|\psi|=|\psi_{max}|$, the cyan dashed line for $|\psi|=|\psi_{max}|/2$ and the magenta asterisks for $|\psi|=|\psi_{max}|/8$. It can be appreciated how the repulsion prevents the cross of solitons at $x=0$.
} %the closer between the solitons.
\label{fig12}
\end{figure}

However, initially out-of-phase solitons, i.e., having initial relative phase $\Delta\phi=\pi$, do not recombine, but move outwards after the collision (see Fig.~{\ref{fig11}}), and  the distance between them grows indefinitely. 
%Thus within the allowed numerical accuracy we established repulsion of the out-of-phase solitons. 
The repulsion between the out-of-phase solitons is additionally illustrated in Fig.~{\ref{fig12}}, which shows how the solitons do not cross at $x=0$. %dependence of the distance between the solitons on the propagation distance. 
%The repulsion between the out-of-phase solitons is additionally illustrated in Fig.~{\ref{fig12}} which shows the $x$-coordinates of the soliton peaks:
%\begin{equation}
%x_{L,R}^{(1)} = \arg_{x}  \max_{x,y} |\psi_{L, R}|,
%\end{equation}
%where $\psi_{L}$ and $\psi_R$ is the left and right solitons, respectively. Figure~{\ref{fig12}} also shows $x$-positions of two other intensity levels defined as
%\begin{eqnarray}
%x_{L}^{(\epsilon)} &=& \max_x \left\{ x \mid  |\psi_L(x, y=0)|=\epsilon\psi_{\max} \right\},\\
%x_{R}^{(\epsilon)} &=& \min_x \left\{ x \mid   |\psi_R(x, y=0)|=\epsilon\psi_{\max} \right\},  
%\end{eqnarray}
%where $\epsilon$ takes values $1/2$ and $1/8$, and $\psi_{\max}$ is the maximal intensity: $\psi_{\max} \max_{x,y}|\psi_{L, R}|$ which is the same for the left and for the right solitons.
% evolution in $z$ of particular positions in tems of the maximum amplitudes of the solitons remove the possibility of a crossing after the interaction.  %where the absolute value of $\psi_{1,2,3}$ takes a fixed quantity,

\section{Conclusions}
\label{sec:concl}

We have studied a model of a triple-core wave guide described by a system of three coupled two-dimensional nonlinear Schr\"odinger equations with the  cubic (focusing) -- quintic (defocusing) nonlinearity. In the first part of the work, we have considered the conservative case and classified possible families of bright solitons. The most interesting effect found is the symmetry-breaking bifurcation occurs at varying strength of the coupling or at growing propagation constant. The stability of the found solitons has been addressed in details.

In the second part of the work, we have extended the analysis onto the $\PT$-symmetric system, where one of the waveguides was lossy and another one active, with gain and losses balancing each other. We have demonstrated that such three-waveguide structure supports solitons. Two branches of solitons can be stable, at least for sufficiently weak gain-and-losses. Unlike in the case of two coupled equations \cite{Driben,Malomedgainloss}, the branches of $\PT$-symmetric solitons can be continued into the domain of arbitrarily strong gain-and-losses parameter. However, in this last limit the $\PT$-symmetry is broken what means instability of the solitons. Finally, the interactions and collisions between the solitons were briefly investigated. We observed merge of two in-phase solitons into a stable one and repelling of two out-of-phase solitons.

%The present work paves the way for several potential extensions. One of them is the analysis of 2D %vortex solitons (in a system of two equations such solutions were analyzed in %\cite{Malomedcouplings}). Another interesting issue is a detailed investigation of a system of %three equations assembled in an  \textit{open} array, in contrast with the circular array adopted %in this paper.

% geometry which corresponds
 %The reason is that it is well known that the CQNLSE admits stable solutions with vorticity \cite{Prytula}.
%
%It is noteworthy to mention that an interesting extension of our system is the analysis of 2D vortex solitons. The reason is that it is well known that the CQNLSE admits stable solutions with vorticity \cite{Prytula}. The study of vortex in the two equations model was also reported in \cite{Malomedcouplings}, so it is natural that its study in the system (\ref{3eqsgamma}) is the next step to perform in future works. \\ \\

\acknowledgments
DF thanks \'{A}ngel Paredes for discussions. DF is grateful to the Centro de F\'{i}sica Te\'{o}rica e Computacional of Lisbon University for the hospitality during his stay in Lisbon where part of this work was carried out. The work of DF is supported by the FPU Ph.D. programme, the FPU program of short stays and through grant EM2013/002 of Xunta de Galicia. The work of VVK and DAZ was supported by the FCT (Portugal) through the grants UID/FIS/00618/2013 and PTDC/FIS-OPT/1918/2012.
%-------------------------------------------------------------------------
\\
\appendix
\section{Linear stability eigenvalue problem}
\label{appA}

The substitution of (\ref{3pert}) in (\ref{3eqsgamma}) and the subsequent linearization with respect to $U_\pm$, $V_\pm$ and $W_\pm$ leads to the following eigenvalue problem  (we additionally assume $\gamma=0$):
\begin{eqnarray}
i \sigma U_+ + LU_+  + 2 |U|^2 U_+  \nonumber \quad \quad \quad \quad   \label{3pertsolv}  \\ + U^2 U_- - 3 |U|^4 U_+ - 2 U^3 U^* U_- + \alpha V_+ + \beta W_+=0;   \quad \quad \quad \quad  \nonumber \\
- i \sigma U_- + LU_- + 2 |U|^2 U_-  \nonumber \quad \quad \quad \quad   \\ + U^{2*} U_+ - 3 |U|^4 U_- - 2 U^{3*} U U_+ + \alpha V_- + \beta W_-=0;  \quad \quad \quad \ \   \nonumber  \\
i \sigma V_+ LV_+ + 2 |V|^2 V_+  \nonumber \quad \quad \quad \quad \ \  \\ +  V^2 V_- - 3 |V|^4 V_+ - 2 V^3 V^* V_- + \alpha U_+ + \alpha W_+=0;\quad \quad \quad \quad \\
- i \sigma V_- + LV_-  + 2 |V|^2 V_-  \nonumber \quad \quad \quad \quad \ \  \\ + V^{2*} V_+ - 3 |V|^4 V_- - 2 V^{3*} V V_+ + \alpha U_- + \alpha W_-=0;  \quad \quad \quad \ \ \nonumber \\
i \sigma W_+ + LW_+ + 2 |W|^2 W_+  \nonumber \quad \quad  \  \\ + W^2 W_- - 3 |W|^4 W_+ - 2 W^3 W^* W_- + \alpha V_+ + \beta U_+=0;  \quad \quad \  \nonumber  \\
 - i \sigma W_- + LW_- + 2 |W|^2 W_-  \nonumber \quad \quad  \ \\ + W^{2*} W_+ - 3 |W|^4 W_- - 2 W^{3*} W W_+ + \alpha V_- + \beta U_-=0.  \nonumber \quad \quad \
\end{eqnarray}
Here the linear operator $L$ is defined as
\begin{equation}
L = \frac{d^2}{dr^2} + \frac{1}{r} \frac{d}{dr} -  \frac{n^2}{r^2}-k,
\end{equation}
$\sigma$ is the eigenvalue which characterizes the instability rate, and $n=0,1,2,\ldots$ is the azimuthal index of the perturbation.

%This system can be solved with different numerical methods, for example, using a finite differences scheme. For fixed values of the parameters $k$, $\alpha$ and $\beta$ the solution of (\ref{3pertsolv}) provides the eigenvalues $\sigma$ for that particular case. This allows to know the degree of stability of the soliton calculating the real part of the maximum unstable eigenvalue $S=max(Re(\sigma))$. I

%--------------------------------------------------------------------------

\end{document}